\documentclass[conference]{IEEEtran}
\IEEEoverridecommandlockouts

\usepackage{cite}
\usepackage{amsmath,amssymb,amsfonts}
\usepackage{algorithmic}
\usepackage{graphicx}
\usepackage{subcaption}
\usepackage{textcomp}
\usepackage{xcolor}
\def\BibTeX{{\rm B\kern-.05em{\sc i\kern-.025em b}\kern-.08em
    T\kern-.1667em\lower.7ex\hbox{E}\kern-.125emX}}
\usepackage{cite}
\usepackage{amsmath,amssymb,amsfonts}

\usepackage{graphicx}
\usepackage{textcomp}
\usepackage{amsmath}
\usepackage[numbers]{natbib} 
\usepackage{bbm}
\usepackage{xcolor}
\usepackage[ruled,vlined]{algorithm2e}

\def\minwrt[#1]{\underset{#1}{\mathrm{minimize }}}
\newcommand{\abs}[1]{\left|#1\right|}
\newcommand{\bC}{\mathbf{C}}
\newcommand{\bM}{\mathbf{M}}
\newcommand{\bS}{\mathbf{S}}
\newcommand{\ba}{\mathbf{a}}
\newcommand{\bb}{\mathbf{b}}

\newcommand{\RR}{\mathbb{R}}

\newcommand{\specrec}{\mathbf{R}}
\newcommand{\specsource}{\mathbf{S}}
\newcommand{\specnoise}{\mathbf{W}}

\newcommand{\onevec}{\mathbf{1}}

\begin{document}

\title{Joint Spectrogram Separation and \\TDOA Estimation using Optimal Transport
\thanks{This work is supported by the Finnish Doctoral Program Network in Artificial Intelligence (AI-DOC), grant ID: VN/3137/2024-OKM-6.}
}

\author{\IEEEauthorblockN{Linda Fabiani\IEEEauthorrefmark{1}, Sebastian J. Schlecht\IEEEauthorrefmark{2}, Isabel Haasler\IEEEauthorrefmark{3}, Filip Elvander\IEEEauthorrefmark{1}}
\IEEEauthorblockA{\IEEEauthorrefmark{1}Dept. of Information and Communications Engineering, Aalto University, Finland\\
}
\IEEEauthorblockA{\IEEEauthorrefmark{2}Dept. of Electrical and Computer Engineering, Friedrich-Alexander-Universität Erlangen-Nürnberg, Germany
}

\IEEEauthorblockA{\IEEEauthorrefmark{3}Dept. of Information Technology, Uppsala University, Sweden
}

  }%

\maketitle

\begin{abstract}
Separating sources is a common challenge in applications such as speech enhancement and telecommunications, where distinguishing between overlapping sounds helps reduce interference and improve signal quality. Additionally, in multichannel systems, correct calibration and synchronization are essential to separate and locate source signals accurately. This work introduces a method for blind source separation and estimation of the Time Difference of Arrival (TDOA) of signals in the time-frequency domain. Our proposed method effectively separates signal mixtures into their original source spectrograms while simultaneously estimating the relative delays between receivers, using Optimal Transport (OT) theory. By exploiting the structure of the OT problem, we combine the separation and delay estimation processes into a unified framework, optimizing the system through a block coordinate descent algorithm. We analyze the performance of the OT-based estimator under various noise conditions and compare it with conventional TDOA and source separation methods. Numerical simulation results demonstrate that our proposed approach can achieve a significant level of accuracy across diverse noise scenarios for physical speech signals in both TDOA and source separation tasks.

\end{abstract}

\begin{IEEEkeywords}
Optimal transport, Time Difference Of Arrival, Blind Source Separation, spectrogram
\end{IEEEkeywords}

\section{Introduction}

\begin{figure}[t] 
    \centering
    \begin{subfigure}[b]{0.49\columnwidth}
        \centering
        \includegraphics[width=\linewidth]{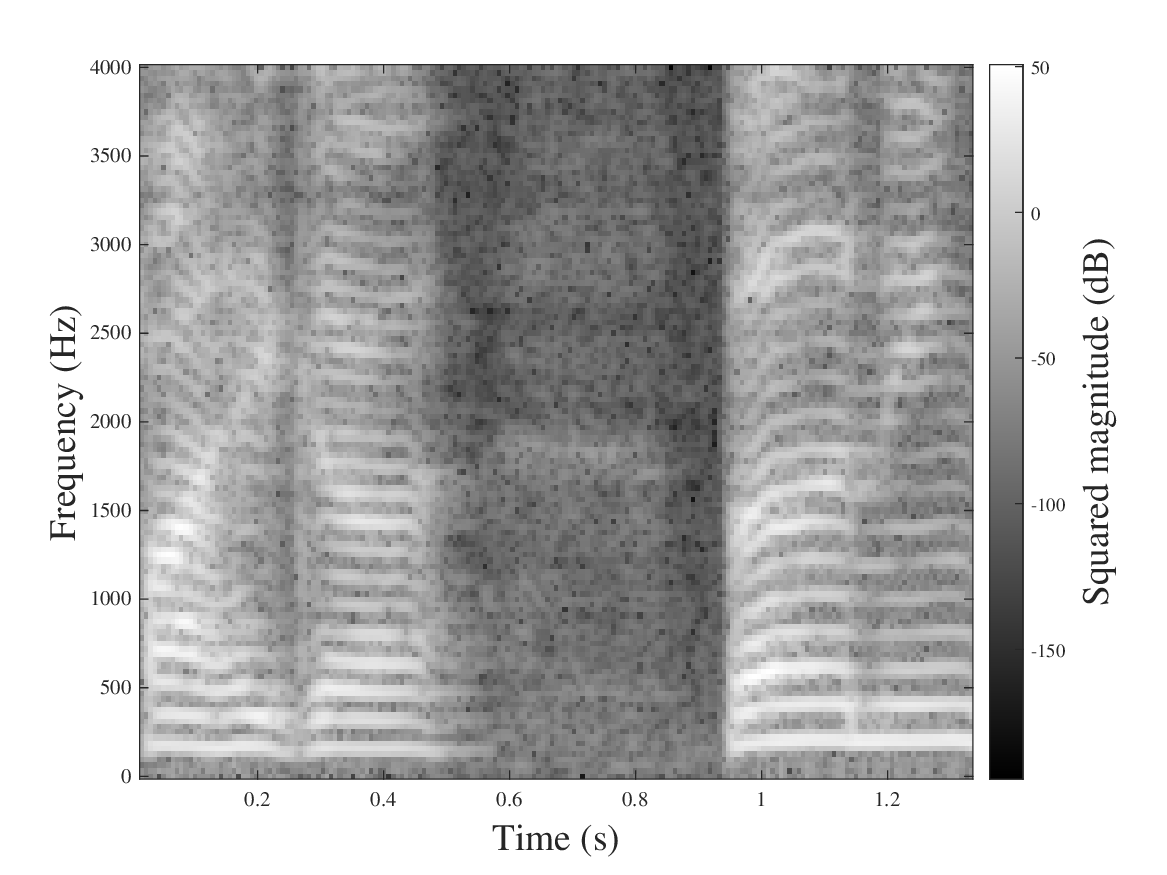}
        \caption{}
        \label{subfig:femaleSpeech}
    \end{subfigure}
    \hfill
    \begin{subfigure}[b]{0.49\columnwidth}
        \centering
        \includegraphics[width=\linewidth]{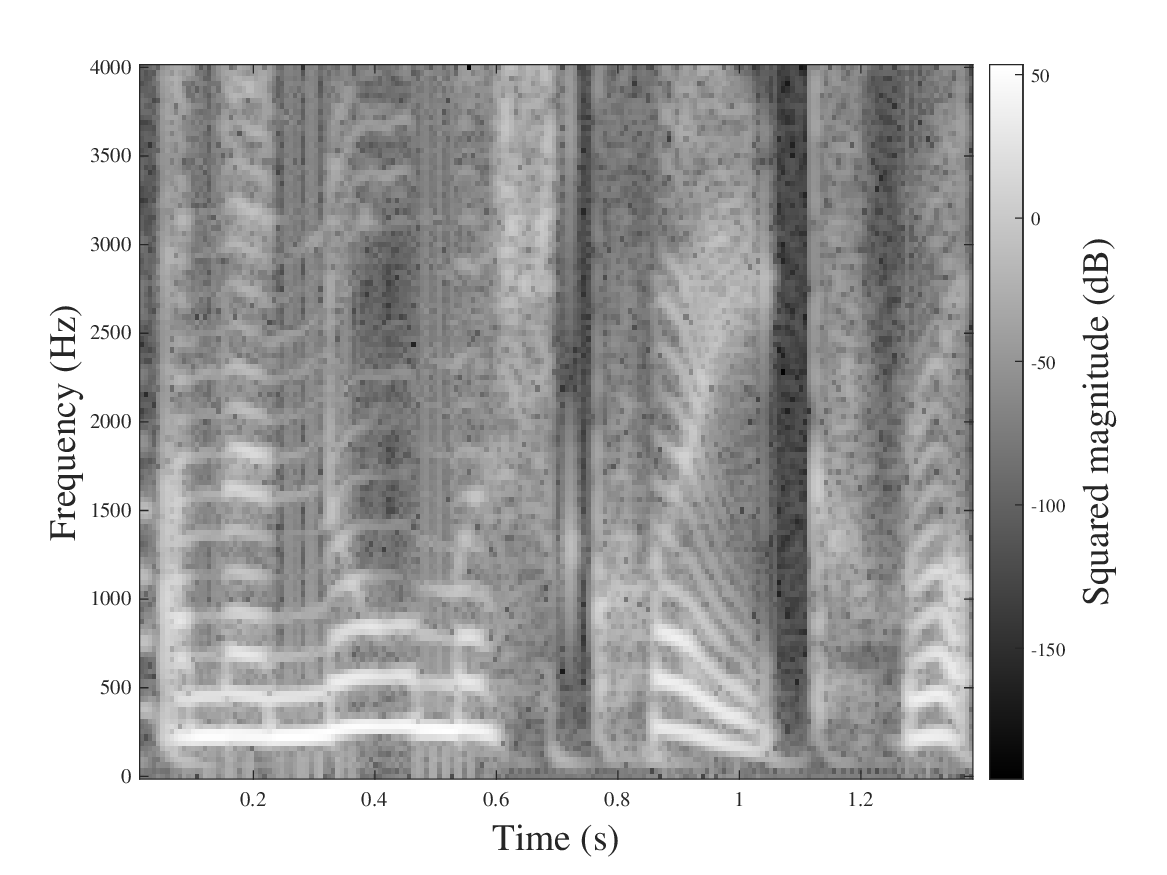}
        \caption{}
        \label{subfig:maleSpeech}
    \end{subfigure}

    \begin{subfigure}[b]{0.49\columnwidth}
        \centering
        \includegraphics[width=\linewidth]{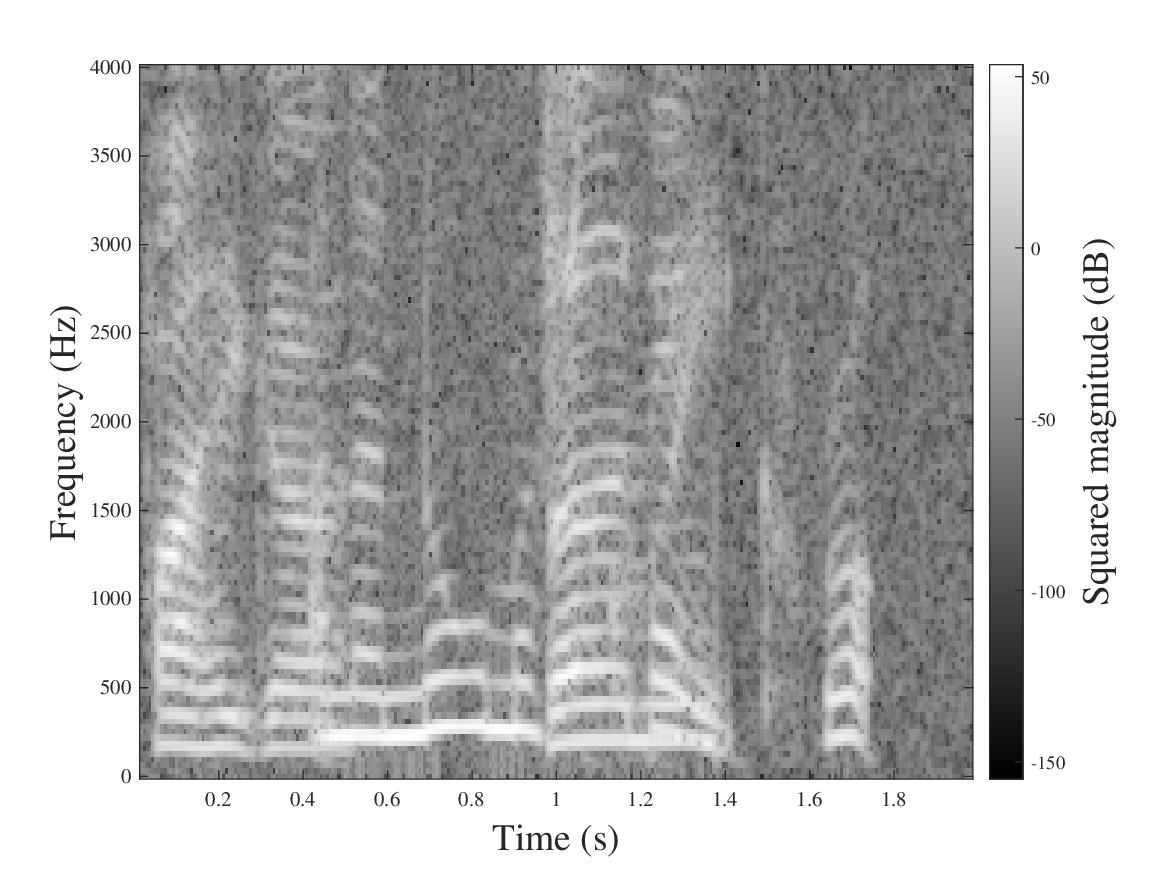}
        \caption{}
        \label{subfig:RX1}
    \end{subfigure}
    \hfill
    \begin{subfigure}[b]{0.49\columnwidth}
        \centering
        \includegraphics[width=\linewidth]{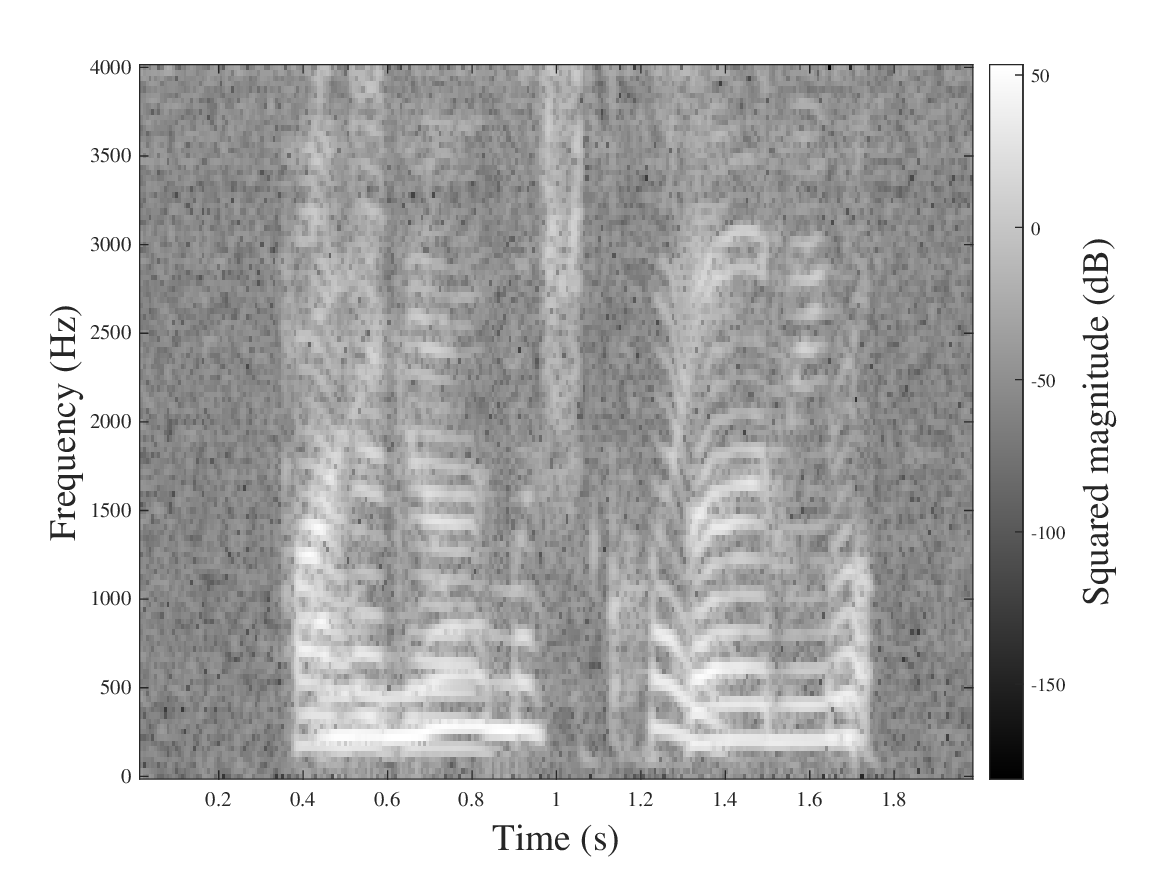}
        \caption{}
        \label{subfig:RX2}
    \end{subfigure}

    \caption{Illustration of source power spectrograms and their corresponding mixtures observed at two receivers. \textbf{Top:} Original female (a) and (b) male speech source signal. \textbf{Bottom:} Observed mixtures at receiver 1 (c) and receiver 2 (d), with distinct propagation delays. The receiver signals include delayed versions of the original source signals and have been zero-padded to a uniform duration of 2 seconds.}
    \label{fig:showDelayedSpectrograms}
\end{figure}

Source separation is a fundamental task used in signal processing, present in many applications such as speech enhancement, noise reduction and telecommunications. However, de-mixing spectral signals in real-world scenarios becomes challenging due to reverberation and 
background noise interference, especially without direct access to the source signals. This absence of ground truth references adds uncertainty, requiring the system to rely on statistical and spatial cues to effectively distinguish and extract individual sources. To address this problem, various Blind Source Separation (BSS) methods have been developed \cite{bssBook}, including Independent Component Analysis (ICA) \cite{ICA} and Non-Negative Matrix Factorization (NMF) \cite{MNMF, shiftNMF}, as well as recent deep learning-based techniques trained on specific data for improved performance \cite{SSDL, wang2018supervisedspeechseparationbased}. Despite their wide usage, these methods can struggle in noisy acoustic environments, or when sources are not instantaneous. Indeed, in multichannel systems, where sensors receive mixtures of signals with varying propagation paths, a fundamental challenge lies in accurately finding the time delays between signals arriving at different locations. For example, Figure \ref{fig:showDelayedSpectrograms} shows two source signals and their corresponding mixtures, illustrating how the two speech signals arrive at the two receivers at different times.
 For this, Time Difference of Arrival (TDOA) estimation plays a crucial role in improving separation, particularly in microphone arrays and spatial filtering applications \cite{TDOAInterp, tdoaMic}. Traditional cross-correlation methods, such as Generalized Cross-Correlation (GCC) and Generalized Cross-Correlation with Phase Transform (GCC-PHAT) \cite{GCC}, are commonly used for delay estimation, even though they also suffer from reduced accuracy in adverse noise conditions \cite{TDOAInterp, Chen2006TimeDE}. 

In this work, we introduce a novel approach to multichannel blind source separation of time-frequency distributions, using the theory of optimal transport (OT) (see, e.g. \cite{TopicsVillani}). OT is a mathematical concept that allows for comparing distributions of mass and has found use in fields including signal processing \cite{Georgiou_axiomatic,multimarginalElvander} and control \cite{Chen_OTcontrol,Lamoline2024GeneRN}.
Recently, the work \cite{ElvanderH25_unmix_dyn_arxiv} has introduced an OT framework for identifying and separating ensembles based on population-level observations and for estimating the dynamics governing these ensembles. In this paper, we cast the BSS problem within this framework, letting the ensembles and dynamics correspond to the (unknown) source signal spectra and inter-microphone time delays, respectively. In particular, by exploiting the temporal delay information between the channels, we formulate an optimal transport problem, where each source is associated with a distinct transport plan that governs the separation of the spectral components. We utilize a block-coordinate descent algorithm that iteratively updates the optimized variables to compute our estimator. As a potential application of our framework, we provide an example of reconstruction of the source signals in the time domain, processing the estimated spectrograms via multichannel Wiener filtering. We demonstrate the robustness of the proposed method in the case of diverse sensor noise and compare it with commonly used methods for TDOA estimation and source separation.

\section{Signal Model}\label{section:signalModel}
Consider a system with $K \in \mathbb{N}$ acoustic sources impinging on $L \in \mathbb{N}$ microphones. The spatial locations of the sources as well as the source signals are unknown.
Let $S_k(f,t)$ and $R_\ell(f,t)$ denote the short-time Fourier transform (STFT) representations of the $k$th source signal and $\ell$th microphone signal, respectively, where $f$ and $t$ are the frequency and time index of a time-frequency bin. Then, using the first microphone as reference, each microphone signal can be written as\footnote{To simplify the exposition, we here assume that differences in signal attenuation between microphones are small, corresponding to the microphones being closely spaced.}
\begin{align}\label{eq:stft_domain_sensors}
    R_\ell(f,t) = \sum_{k=1}^K S_k(f,t-\tau_k^{(\ell)}) + W_\ell(f,t),
\end{align}
where $\tau_k^{(\ell)}$ denotes the delay of the $k$th source to the $\ell$th receiver relative to the reference microphone. That is, $\tau_k^{(\ell)}$ is the TDOA of the $k$th source for the microphone pair with indices $(1,\ell)$. Note that $\tau_k^{(1)} = 0$, $k = 1,\ldots,K$. Furthermore, $W_\ell(t)$ denotes the STFT of the sensor noise, which we herein assume to be well-modelled as spatially and temporally white Gaussian noise with variance $\sigma^2$. Herein, we aim to separate the different signal sources and estimate their TDOAs based on spectral representations. To that end, let $\specrec^{(\ell)} \in \RR_+^{T\times F}$ be the spectrogram of the sensor signal in \eqref{eq:stft_domain_sensors}, defined as the squared magnitude of the STFT, represented as a matrix, i.e., $[\specrec^{(\ell)}]_{t,f} = \abs{R_\ell(f,t)}^2$  where $T$ and $F$ are the number of time and frequency bins, respectively. Correspondingly, let $\specsource_k^{(\ell)} \in \RR_+^{T \times F}$ be the spectrogram of the $k$th source in the $\ell$th microphone, and let $\specnoise_k^{(\ell)}$ be the spectrogram of the sensor noise. Note here that $\specsource_k^{(\ell)}$ can be obtained from $\specsource_k^{(1)}$ by shifting its rows in accordance with $\tau_k^{(\ell)}$. Lastly, assuming that the different source signals are statistically independent, we approximate the microphone spectrograms $\specrec^{(\ell)}$ as additive mixtures of the source spectrograms\footnote{In our experiments, we observe that our proposed method performs well despite this approximation.} \cite{LiutkusAdditivity, genWienerAdditivity}:
\begin{align}\label{eq:spectrogram_sensors}
    \specrec^{(\ell)} \approx \sum_{k = 1}^K \specsource_k^{(\ell)} + \specnoise^{(\ell)}.
\end{align}

With this, our aim is to estimate the source spectrograms\footnote{Without loss of generality, we define the source spectrogram for source $k$ as $\mathbf{S}_k^{(1)}$, i.e., as it would appear in the reference microphone in noise-free single-source settings.} $\specsource_k^{(1)}$ and TDOAs $\tau_k^{(\ell)}$ given only the microphone spectrograms $\specrec^{(\ell)}$. In particular, we propose to view the spectrograms as mass distributions on the time-frequency plane, allowing us to cast the problem in the framework of optimal transport.

\section{Optimal Transport}\label{section:OT}

Consider two discrete distributions, represented with vectors $\ba \in \mathbb{R}^N$ and $\bb \in \mathbb{R}^N$
for some $N \in \mathbb{N}$.
The (discrete) Monge-Kantorovich problem of OT is
\begin{equation} \label{eq:classic_ot_problem}
    \begin{aligned}
     \minwrt[\bM  \in \mathbb{R}_{+}^{N\times N}]&\quad \langle \bC,\bM\rangle 
        \\\text{s.t. }&\quad \bM\onevec_N = \ba \;,\; \bM^T\onevec_N = \bb.
    \end{aligned}
\end{equation}
%
%
Here, $\bC \in \mathbb{R}_{+}^{N\times N}$ is the cost matrix, where each entry $[\bC]_{ij}$ represents the cost of transporting a unit of mass from the 
$i$-th point in the source distribution to the 
$j$-th point in the target distribution, and $\onevec_N\in \mathbb{R}^N$ is a vector of ones. The matrix $\bM$ represents the transport plan, in which each element $[\bM]_{ij}$  describes how much mass is transported from one distribution to another \cite{TopicsVillani, peyré2020computational}.
Thus, the constraints in \eqref{eq:classic_ot_problem} impose that the transport plan $\bM$ transports the mass from $\ba$ to $\bb$.

\section{Method Outline}\label{section:methodOutline}

Herein, we propose to jointly solve the blind source separation
and TDOA estimation tasks using an OT formulation building on the framework in \cite{ElvanderH25_unmix_dyn_arxiv}. More precisely, based on estimates of the receiver's spectrograms $\specrec^{(\ell)}$ 
we identify the source spectrograms $\specsource_k^{(1)}$ and the TDOAs $\tau_k^{(\ell)}$ in \eqref{eq:spectrogram_sensors}. Let $\mathbf{r}_f^{(\ell)}$ and $\mathbf{s}_{k,f}^{(\ell)}$ denote the $f$th column of the matrices $\specrec^{(\ell)}$ and $\specsource_k^{(\ell)}$, respectively. 
We describe the distribution shift due to the TDOA, that is, from $\mathbf{s}_{k,f}^{(1)}$ to $\mathbf{s}_{k,f}^{(\ell)}$ by a transport plan $\bM^{f, \ell}_k$, for $k=1,\dots,K$, $f=1,\dots,F$, and $\ell=2,\dots,L$.
These transport plans are feasible if they satisfy $\bM^{f,\ell}_k \mathbf{1} = \mathbf{s}_{k,f}^{(1)}$ and $ (\bM^{f,\ell}_k)^T \mathbf{1} = \mathbf{s}_{k,f}^{(\ell)}$ for $\ell=2,\dots,L$.
Given a TDOA $\tau_k$ we measure the transportation cost of a transport plan $\bM^{f,\ell}_k$ as $ \langle \bC(\tau_k^{(\ell)}), \bM^{f,\ell}_k \rangle$, where
\begin{equation}\label{eq:costwithtau}
    \begin{aligned}
        \left[\bC(\tau)\right]_{ij} =  \left(  (t_i- t_j) - \tau \right)^2.
    \end{aligned}
\end{equation}

Thus, our goal is to accurately separate the source spectrograms $\bS^{(1)}_k$ by exploiting the structure of the optimal transport problem. Each source is associated with a distinct transport plan $\bM_k^{f,\ell}$, which enables the decomposition of the total objective function into individual contributions for each source.
%
%
Since the true time delays are unknown, they must also be optimized along with the transport plans. 
To sum up, we find the source spectrograms and TDOAs by solving the optimal transport type problem 
\begin{equation}
    \begin{aligned}
        \minwrt[\bM^{f,\ell}_k,\tau_k^{(\ell)},\mathbf{s}_{k,f}^{(\ell)}] \ & \ \sum_{\ell=2}^L \sum_{k=1}^{K}\sum_{f=1}^{F} \langle \bC(\tau_k^{(\ell)}), \bM^{f,\ell}_k \rangle
		\\ \text{subject to } \ & \ \bM^{f,\ell}_k \mathbf{1} = \mathbf{s}_{k,f}^{(1)}, \ \text{for } \ell=2,\dots,L \\
        & (\bM^{f,\ell}_k)^T \mathbf{1} = \mathbf{s}_{k,f}^{(\ell)}, \ \text{for } \ell=2,\dots,L \\
        & \sum_{k=1}^{K} \mathbf{s}_{k,f}^{(\ell)}  =  \mathbf{r}_f^{(\ell)}, \quad \text{for } \ell=1, ...,L \\
        &\text{for } k=1,\dots,K,\ f=1,\dots,F.
     \end{aligned}
     \label{eq:fullOTproblem}
\end{equation}

The constraints in \eqref{eq:fullOTproblem} enforce that the spectrogram mass of each source is fully separated from the mixture and distributed in the relative transport plan. Additionally, the third constraint imposes mass conservation across the system, meaning that the sum of the estimated source spectrograms at each sensor must exactly match that of the sensor spectrograms. In the noise-free case, it may be noted that if the TDOAs in \eqref{eq:fullOTproblem}  are selected as the ground truth values, then the objective value can be set to zero by transport plans that separate the sources. For a discussion on this, see \cite[Section II.B]{ElvanderH25_unmix_dyn_arxiv}.

The optimization problem in \eqref{eq:fullOTproblem} is bi-convex. For fixed time-delays it is linear in the transport plans and source spectra, and due to the choice of \eqref{eq:costwithtau}, the program is convex in the time-delays for fixed transport plans and source spectra. To solve \eqref{eq:fullOTproblem}, we employ a block-coordinate descent algorithm, alternating between minimizing \eqref{eq:fullOTproblem} with respect to the set of transport plans and source spectrograms, and with respect to the time-delays $\tau_k^{(\ell)}$. It may be noted that, due to \eqref{eq:costwithtau}, the minimization with respect to the $\tau_k^{(\ell)}$ corresponds to solving an unconstrained quadratic program, which can be done analytically. The procedure is summarized in Algorithm~\ref{alg:BCD}. Furthermore, we note that the problem in \eqref{eq:fullOTproblem} satisfies the conditions of \cite[Proposition~2]{ElvanderH25_unmix_dyn_arxiv}, and Algorithm~\ref{alg:BCD} is therefore guaranteed to converge to a (local) minimum.
Given a solution to \eqref{eq:fullOTproblem}, the source spectrograms $\specsource_k^{(1)}$, $k = 1,\ldots,K$, can be reconstructed from  $\mathbf{s}_{k,f}^{(1)}$ as
\begin{align*}
    \hat{\specsource}_k^{(1)} = \begin{bmatrix}
        \mathbf{s}_{k,1}^{(1)} & \mathbf{s}_{k,2}^{(1)} & \ldots & \mathbf{s}_{k,F}^{(1)}
    \end{bmatrix}
\end{align*}
where we remind that $\mathbf{s}_{k,1}^{(1)} = \bM^{f,\ell}_k \mathbf{1}$ for any $\ell \in \{2,\ldots,L\}$.
%

%
%
%
%
%


\begin{algorithm}[t]
    \caption{{Block Coordinate Descent}}\label{alg:BCD}
    \KwData{Initialize  $\tau_{k}^{(\ell)}$ for $k=1,\dots,K$ }
     \While{not converged}{
      minimize (\ref{eq:fullOTproblem}) w.r.t. $\bM_k^{f,\ell}$\;
      minimize (\ref{eq:fullOTproblem}) w.r.t. $\tau_k^{(\ell)}$
     }  
\end{algorithm}

\begin{figure}[t]
\centerline{\includegraphics[width=\linewidth]{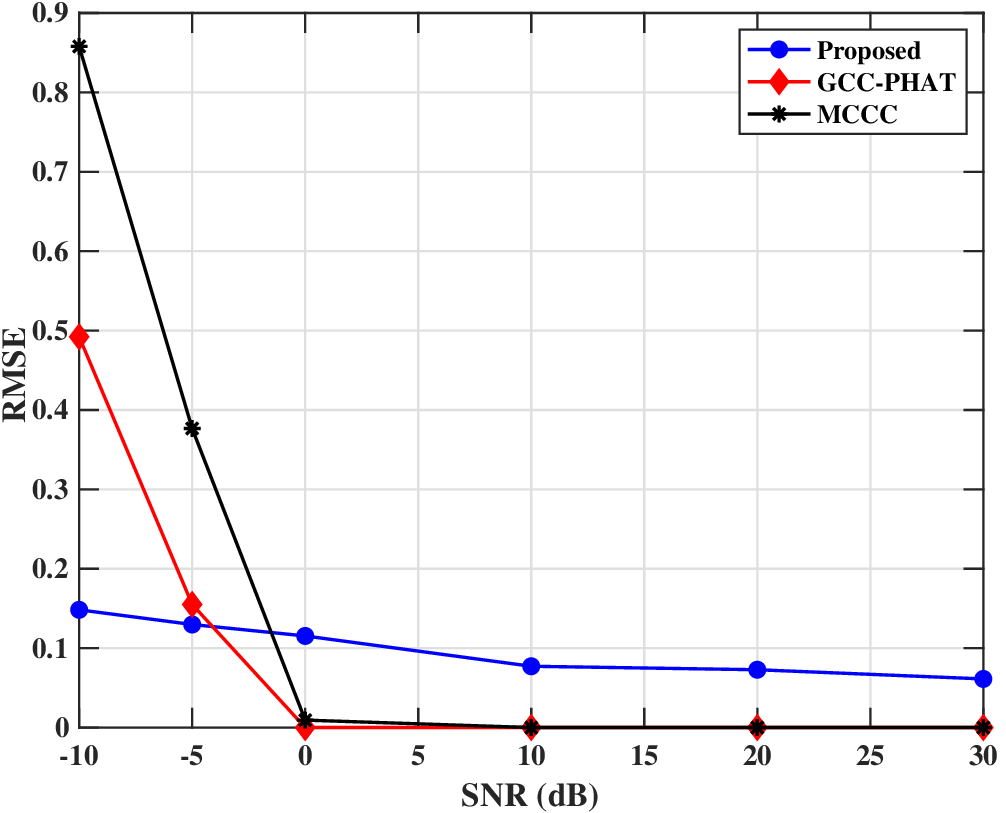}}
\caption{Performance comparison for the TDOA estimation task. The evaluated methods are all assessed in the presence of fractional delays.}
\label{fig:TDOA}
\end{figure}

%

 \section{Numerical Experiments}

 The performance of the proposed method for joint source separation and TDOA estimation is assessed and compared to other commonly used techniques across a series of experiments. The evaluation is performed on a 2$\times$2 source-receiver system, where two distinct speech segments are used as source signals. Specifically, they consist of two phrases, one spoken by a male speaker and the other by a female speaker. To simulate a realistic multichannel recording scenario, the two signals are then artificially overlapped and delayed to create two receiver mixtures, with a total duration of 2 seconds. All signals are first downsampled to a frequency $f_s=8$ kHz and transformed to spectrograms computing the STFT using a Hann window of size of 256, Hop size of 200 and FFT size of 256. An example of the data used is shown in Figure \ref{fig:showDelayedSpectrograms}. For each experiment, we perform 100 Monte Carlo simulations, where the relative delays are randomly initialized from a limited set of values based on the length of the speech signal.  
 
First, we evaluate the performance of the TDOA estimation by comparing our method to standard approaches, including Generalized Cross-Correlation with Phase Transform (GCC-PHAT) \cite{GCC}, and the Multiple Cross-Correlation Coefficient (MCCC) method \cite{Chen2006TimeDE}, as implemented in \cite{onFrequency}. We quantify the error in delay estimation, calculating the root mean squared deviations from the ground truth delays, averaging across all sources as 

\begin{equation}
 \mathrm{RMSE} = \sqrt{\frac{1}{K} \sum_{k=1}^{K} \left( \tau_k - \hat{\tau}_k\right)^2} ,
\end{equation}

for multiple Signal-to-Noise (SNR) scenarios. The results shown in Figure \ref{fig:TDOA} suggest that our proposed approach demonstrates greater robustness at very low SNR conditions,  where conventional cross-correlation techniques struggle due to noise sensitivity. While the OT method exhibits slightly larger RMSE for high SNR levels, it still achieves a strong performance considering the end-to-end complexity and functionality of the full OT estimator. Additionally, it is worth noting that the OT method operates in the time-frequency domain, meaning its accuracy is significantly influenced by the temporal resolution of the STFT and the convergence tolerance of the algorithm. 

\begin{figure}[t]
\centerline{\includegraphics[width=\linewidth]{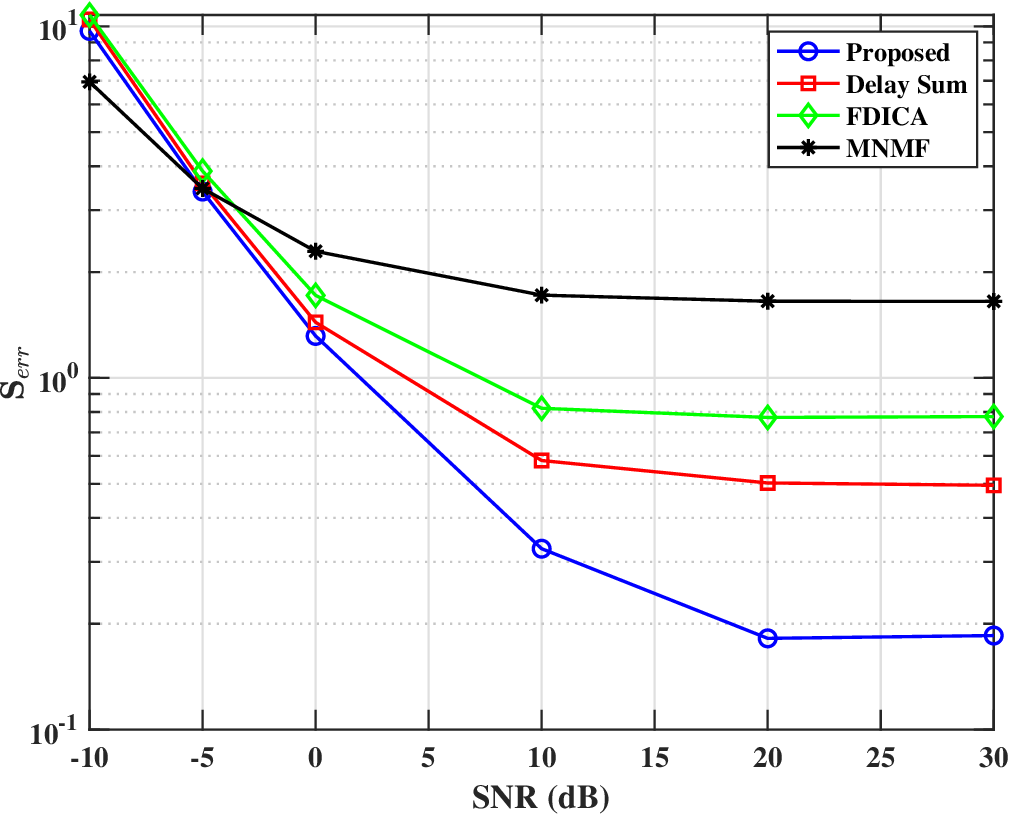}}
\caption{Comparison of spectrogram reconstruction error for different source separation methods.}
\label{fig:specRec}
\end{figure}

The second experiment assesses the accuracy of the reconstructed spectrograms following source separation.  We compare our method against two other BSS techniques: Frequency-Domain Independent Component Analysis (FDICA) \cite{5473129, MURATA20011} and Multichannel Non-Negative Matrix Factorization (MNMF) \cite{Ozerov}. As a baseline, we also include the Delay-and-Sum Beamformer. The reconstruction error between the original and estimated spectrograms is normalized with respect to the reference source and averaged across all sources as 
\begin{equation}\label{eq:spec_rec}
    \bS_{err} = \frac{1}{K}\sum_{k=1}^K \frac{ \| \bS_k - \hat{\bS}_k \|^2_F }{\|\bS_k\|^2_F}  .
\end{equation}

As shown in Figure \ref{fig:specRec}, the proposed OT approach achieves the smoothest and lowest reconstruction error overall. For low SNR values, the results indicate that all methods produce reconstructions that are predominantly affected by noise. Indeed, the optimal transport estimator operates on spectrogram mixtures that contain substantial distortion, so the transport process naturally redistributes the noise across the signals. Without an additional constraint or term in the objective function to mitigate this effect, the noise is inevitably incorporated into the transport plans and, consequently, the reconstructed spectrograms. Future work will address this challenge by exploring unbalanced optimal transport methods \cite{CHIZAT20183090}, which offer a more flexible framework for de-noising tasks.

In the final experiment, we apply our method to reconstruct time-domain signals using the multichannel Wiener filter from \cite{SSDL}. This process is useful for reducing residual noise and allowing a time-domain reconstruction of the estimated power spectrograms. To assess the performance, we measure the improvement in Signal-to-Distortion Ratio (SDR), which quantifies the reduction of interference and noise in the reconstructed signal compared to that of the receiver mixtures,
\begin{equation}
    \Delta \mathrm{SDR} = 10 \log_{10} \left( \frac{\sigma_{s}^2}{\sigma_{D,\hat{s}}^2} \right) - 10 \log_{10} \left( \frac{\sigma_{\text{s}}^2}{\sigma_{D, mix}^2} \right), 
\end{equation}

where $ \sigma_{s}^2 = \mathbb{E} \left[ s_k^2 \right]$  denotes the variance of the original clean source signal,  $\sigma_{D, \hat{s}}^2 = \mathbb{E} \left[ (s_k - \hat{s}_k)^2 \right]$ represents the variance of the distortion in the estimated signal, i.e., the error between the original and reconstructed source, and  $\sigma_{D,mix}^2 = \mathbb{E} \left[ (s_k - s_{mix})^2 \right] $ corresponds to the variance of the distortion in the observed mixture signal before separation.

\begin{figure}[t]
\centerline{\includegraphics[width=\linewidth]{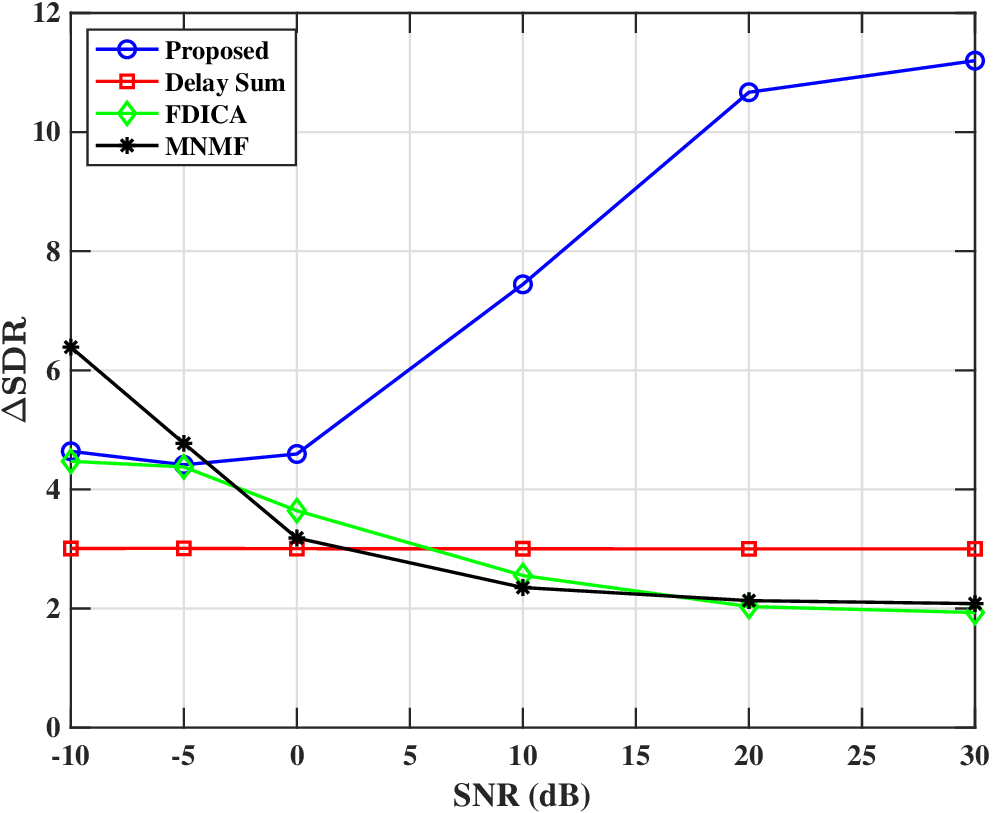}}
\caption{Comparison of Signal-to-Distortion ratio increase for different source separation methods.  }
\label{fig:SDRincrease}
\end{figure}

 The results in Figure \ref{fig:SDRincrease} indicate that the proposed OT method consistently outperforms the other approaches, particularly at higher SNRs, where it achieves a significant SDR increase. In contrast, FDICA and MNMF exhibit a decline in SDR as the SNR increases, indicating that both methods struggle to properly separate the given data, often reporting estimates that are still mixed. 
 Overall, the experimental results validate the effectiveness of the proposed method for joint source separation and TDOA estimation, demonstrating its capability to handle both challenges simultaneously with good accuracy and efficiency.

\section*{Conclusion}
In this paper, we develop a novel approach for the task of source separation and TDOA estimation for a multichannel acoustic system. Our proposed method is based on a multi-step optimization formulated using an optimal transport framework, which is solved efficiently as a block-coordinate descent algorithm. By exploiting the structure of the optimal transport problem, we achieve effective separation of the individual source spectrograms through their transport plans estimates, while precisely evaluating the relative delay values. We utilize these estimates to reconstruct the spectrograms, which are then included in a multichannel Wiener filtering application to recover the time domain signals. The proposed method shows a promising performance in the case of physical speech data, and improved results compared to commonly used delay estimation and source separation methods. This work demonstrates a solid foundation for developing a method for multichannel source separation and TDOA estimation for real room-acoustic reverberant settings and microphone array structures.

\bibliographystyle{ieeetr}
\bibliography{references.bib}

\begin{thebibliography}{10}

\bibitem{bssBook}
P.~Comon and C.~Jutten, {\em Handbook of Blind Source Separation: Independent Component Analysis and Applications}.
\newblock USA: Academic Press, Inc., 1st~ed., 2010.

\bibitem{ICA}
P.~Comon, ``Independent component analysis, a new concept?,'' {\em Signal Processing}, vol.~36, no.~3, pp.~287--314, 1994.
\newblock Higher Order Statistics.

\bibitem{MNMF}
A.~Ozerov and C.~Fevotte, ``Multichannel nonnegative matrix factorization in convolutive mixtures for audio source separation,'' {\em IEEE Transactions on Audio, Speech, and Language Processing}, vol.~18, no.~3, pp.~550--563, 2010.

\bibitem{shiftNMF}
D.~FitzGerald, M.~Cranitch, and E.~Coyle, ``Shifted non-negative matrix factorisation for sound source separation,'' in {\em IEEE/SP 13th Workshop on Statistical Signal Processing, 2005}, pp.~1132--1137, 2005.

\bibitem{SSDL}
A.~A. Nugraha, A.~Liutkus, and E.~Vincent, ``Multichannel music separation with deep neural networks,'' in {\em 2016 24th European Signal Processing Conference (EUSIPCO)}, pp.~1748--1752, 2016.

\bibitem{wang2018supervisedspeechseparationbased}
D.~Wang and J.~Chen, ``Supervised speech separation based on deep learning: An overview,'' 2018.

\bibitem{TDOAInterp}
J.~Benesty, J.~Chen, and Y.~Huang, ``Time-delay estimation via linear interpolation and cross correlation,'' {\em IEEE Transactions on Speech and Audio Processing}, vol.~12, no.~5, pp.~509--519, 2004.

\bibitem{tdoaMic}
J.~Chen, J.~Benesty, and Y.~Huang, ``Robust time delay estimation exploiting redundancy among multiple microphones,'' {\em IEEE Transactions on Speech and Audio Processing}, vol.~11, no.~6, pp.~549--557, 2003.

\bibitem{GCC}
C.~Knapp and G.~Carter, ``The generalized correlation method for estimation of time delay,'' {\em IEEE Transactions on Acoustics, Speech, and Signal Processing}, vol.~24, no.~4, pp.~320--327, 1976.

\bibitem{Chen2006TimeDE}
J.~Chen, J.~Benesty, and Y.~Huang, ``Time delay estimation in room acoustic environments: An overview,'' {\em EURASIP Journal on Advances in Signal Processing}, vol.~2006, pp.~1--19, 2006.

\bibitem{TopicsVillani}
C.~Villani, {\em Topics in optimal transportation}.
\newblock Graduate studies in mathematics ; v. 58, Providence, R.I: American Mathematical Society, 2003.

\bibitem{Georgiou_axiomatic}
T.~T. Georgiou, J.~Karlsson, and M.~S. Takyar, ``Metrics for power spectra: An axiomatic approach,'' {\em IEEE Transactions on Signal Processing}, vol.~57, no.~3, pp.~859--867, 2009.

\bibitem{multimarginalElvander}
F.~Elvander, I.~Haasler, A.~Jakobsson, and J.~Karlsson, ``Multi-marginal optimal transport using partial information with applications in robust localization and sensor fusion,'' {\em Signal Processing}, vol.~171, pp.~1--19, June 2020.

\bibitem{Chen_OTcontrol}
Y.~Chen, T.~T. Georgiou, and M.~Pavon, ``Optimal transport in systems and control,'' {\em Annual Review of Control, Robotics, and Autonomous Systems}, vol.~4, no.~Volume 4, 2021, pp.~89--113, 2021.

\bibitem{Lamoline2024GeneRN}
F.~Lamoline, I.~Haasler, J.~Karlsson, J.~Gonçalves, and A.~Aalto, ``Gene regulatory network inference from single-cell data using optimal transport,'' {\em bioRxiv}, 2024.

\bibitem{ElvanderH25_unmix_dyn_arxiv}
F.~Elvander and I.~Haasler, ``Mixtures of ensembles: {S}ystem separation and identification via optimal transport,'' {\em arXiv:2503.13362}, 2025.

\bibitem{LiutkusAdditivity}
A.~Liutkus, R.~Badeau, and G.~Richard, ``Gaussian processes for underdetermined source separation,'' {\em IEEE Transactions on Signal Processing}, vol.~59, no.~7, pp.~3155--3167, 2011.

\bibitem{genWienerAdditivity}
A.~Liutkus and R.~Badeau, ``Generalized wiener filtering with fractional power spectrograms,'' in {\em 2015 IEEE International Conference on Acoustics, Speech and Signal Processing (ICASSP)}, pp.~266--270, 2015.

\bibitem{peyré2020computational}
G.~Peyré and M.~Cuturi, ``Computational optimal transport: With applications to data science,'' {\em Foundations and Trends® in Machine Learning}, vol.~11, no.~5–6, p.~355–607, 2019.

\bibitem{onFrequency}
J.~R. Jensen, J.~K. Nielsen, M.~G. Christensen, and S.~Holdt~Jensen, ``On frequency domain models for tdoa estimation,'' in {\em 2015 IEEE International Conference on Acoustics, Speech and Signal Processing (ICASSP)}, pp.~11--15, 2015.

\bibitem{5473129}
H.~Sawada, S.~Araki, and S.~Makino, ``Underdetermined convolutive blind source separation via frequency bin-wise clustering and permutation alignment,'' {\em IEEE Transactions on Audio, Speech, and Language Processing}, vol.~19, no.~3, pp.~516--527, 2011.

\bibitem{MURATA20011}
N.~Murata, S.~Ikeda, and A.~Ziehe, ``An approach to blind source separation based on temporal structure of speech signals,'' {\em Neurocomputing}, vol.~41, no.~1, pp.~1--24, 2001.

\bibitem{Ozerov}
A.~Ozerov and C.~Fevotte, ``Multichannel nonnegative matrix factorization in convolutive mixtures for audio source separation,'' {\em IEEE Transactions on Audio, Speech, and Language Processing}, vol.~18, no.~3, pp.~550--563, 2010.

\bibitem{CHIZAT20183090}
L.~Chizat, G.~Peyré, B.~Schmitzer, and F.-X. Vialard, ``Unbalanced optimal transport: Dynamic and kantorovich formulations,'' {\em Journal of Functional Analysis}, vol.~274, no.~11, pp.~3090--3123, 2018.

\end{thebibliography}

\end{document}